\documentclass[amsmath,amssymb,groupedaddress]{revtex4}
\usepackage{graphicx}
\usepackage{epstopdf}
\usepackage{xcolor}
\usepackage{amsmath,amsthm,amssymb}
\newcommand{\be}{\begin{equation}}
\newcommand{\ee}{\end{equation}}
\newcommand{\ba}{\begin{eqnarray}}
\newcommand{\ea}{\end{eqnarray}}
\newcommand{\baa}{\begin{eqnarray}}
\newcommand{\eaa}{\end{eqnarray}}
\newcommand{\ed}{\end{document}}
\newcommand{\lab}[1]{\label{#1}}
\newcommand{\re}[1]{(\ref{#1})}
\newcommand{\ci}[1]{\cite{#1}}

\begin{document}
\title[Fast forward problem for adiabatic quantum dynamics]{Fast forward problem for adiabatic quantum dynamics:\\
Estimation of the energy cost}
\author{J.D.~Matrasulov$^{1,2}$, J.R.~Yusupov$^{3,1}$ and Kh.Sh.~Matyokubov$^{4,5}$}
\affiliation{${^a}$
$^1$ Faculty of Physics, National University of Uzbekistan, Vuzgorodok, Tashkent 100174, Uzbekistan\\
$^2$ Institute of Fundamental and Applied Research, National Research University TIIAME, Kori Niyoziy 39, Tashkent 100000, Uzbekistan\\
$^3$ Kimyo International University in Tashkent,\\ 156 Usman Nasyr Str., 100121, Tashkent, Uzbekistan\\
$^4$ Urgench State Pedagogical Institute, 1 Gurlan str., 220100, Urgench, Uzbekistan\\
$^5$ Urgench State University, 14 H. Olimjon Str., 220100, Urgench, Uzbekistan}
\begin{abstract}
 We consider the problem  of energy cost needed for acceleration (deceleration) of the evolution of a quantum system using the Masuda-Nakamura's fast forward protocol. In particular, we focus on dynamics by considering models for a quantum box with a moving wall and harmonic oscillator with time-dependent frequency. For both models we computed the energy needed for acceleration (deceleration) as a function of time. The results obtained are compared with those of other acceleration (deceleration) protocols.
\end{abstract}

\maketitle
\section{Introduction}
Controlling of evolution and manipulation of physical processes in quantum systems is of practical importance in emerging quantum technologies. Achieving such a goal allows to improve device miniaturization and operational speed in nanoscale and quantum functional materials. Solving such a problem requires developing of cost-efficient methods for acceleration and deceleration of quantum processes. One of the methods allowing to accelerate (decelerate) of quantum evolution, including that in adiabatic regime has been proposed in the Ref.~\ci{mas1}, which was later modified for different cases (see Refs.~\ci{mas2,MN11,rv2-kn,iwan,naka,JM,Thermo,heat}) and situations. According to the protocol of this approach evolution of a given quantum system can be accelerated (decelerated) by acting on it by some external electromagnetic potential. The method was called later Masuda-Nakamura's fast forward protocol in quantum mechanics Ref.~\ci{mas1}. It is important to note that there are different protocols for acceleration (deceleration) of quantum evolution. Therefore, an important problem that arises in the context of their practical applications is energy-efficiency of a fast-forward protocol. Such efficiency can be estimated in terms of so called fast forward energy cost which was introduced first in the Ref.~\ci{cost1} and applied later to different physical systems in \ci{cost2}. Here we estimate such a cost for Masuda-Nakamura's fast forward protocol in case of dynamical quantum confinement by considering harmonic oscillator with time-dependent frequency and a quantum box with a moving wall. 

Remarkable feature of  the Masuda-Nakamura's fast forward protocol developed in \ci{mas1} is the fact that it allows to accelerate the time evolution of a quantum system by tuning the external potential that can be reduced to regulation of an additional phase in the wave-function. Much improved version of the prescription was was proposed later in \cite{rv2-kn} which is used for acceleration of the soliton dynamics described in terms of the  nonlinear Schr\"odinger equation  and  tunneling in quantum regime \cite{rv2-kn}.
Different modifications and application of the protocol have been presented later in the Refs.~\ci{rv2-kn,iwan,naka,JM,Thermo,JM}.  One of the advantages of the Masuda-Nakamura protocol is its effective application to so-called adiabatic quantum processes. These latter are the processes occurring in very slowly evolving quantum systems. An interesting problem in this case is so-called  quantum short cuts, implying the shortest path (in time) to the end of the (adiabatic) processes among others. In  \ci{mas2, MN11} the Masuda-Nakamura protocol was adopted to the problem of short cuts. In the literature the problem called "short cuts to adiabaticity"
\ci{Muga,Muga1,Rice,Inver} (called also "transitionless quantum driving" by Berry \ci{Berry}). One should also note successful application of the fast forward protocol to the problem of stochastic \cite{Thermo} and classical \cite{jar} heat engine.  

This paper is organized as follows. In the next section we will present brief description of the fast-forward protocol, following the Ref~\ci{mas1}. Section~3 presents an application of the fast forward protocol to harmonic oscillator with time-dependent frequency and quantum box with moving wall. In Section~4 we calculate energetic cost of fast forward protocol for the harmonic oscillator with time-dependent frequency and compared it with the cost of inverse engineering protocol. In addition, Section~4 presents the results for quantum box with moving boundary conditions. Finally, the section~5 presents some concluding remarks. 

\section{Fast forward protocol for adiabatic quantum dynamics}\label{sec2}
Here we briefly recall the fast forward prescription for adiabatic quantum dynamics and it's application to the harmonic oscillator with time-dependent frequency \ci{JM}. Consider the dynamics of a wavefunction $\psi_0$ under the dynamical confinement $V_0=V_0(x,R(t))$ which varies adiabatically (slowly). This adiabatic change characterized by slowly-varying control parameter $R=R(t)$ which is given by
\begin{eqnarray}
\label{2.1}\label{adia-R}
R(t)=R_{0}+\epsilon t,
\end{eqnarray}
with the growth rate $\epsilon\ll1$. The time-dependent 1D Schr\"{o}dinger equation (1D TDSE) for $\psi_0$ is given as:
\begin{eqnarray}
\label{2.2}
\textit{i}\hbar\frac{\partial\psi_{0}}{\partial t}=-\frac{\hbar^{2}}{2m}\partial_x^{2}\psi_{0}+V_{0}(x,R(t))\psi_{0},
\end{eqnarray}
where the coupling with electromagnetic field is assumed to be absent.
If $R=const$ the problem reduces to an eigenvalue problem for stationary bound state $\phi_0$ which satisfies the time-independent Schr\"{o}dinger equation
\begin{eqnarray}
\label{2.3}
E\phi_{0}=\hat{H}_0\phi_{0}\equiv\left[-\frac{\hbar^{2}}{2m}\partial_x^{2}+V_{0}(x,R)\right]\phi_{0}.
\end{eqnarray}
With use of the eigenstate $\phi_0=\phi_0(x, R)$,
one might conceive the corresponding time-dependent state to be a product of $\phi_0$ and a dynamical factor as,
\begin{eqnarray}\label{2.8}
\phi_{0}(x,R(t))=\bar{\phi}_{0}(x,R(t))e^{\textit{i}\eta(x,R(t))},
\end{eqnarray}
As it stands, however, $\psi_{0}$ does not satisfy TDSE in Eq.~(\ref{2.2}). To overcome this difficulty we introduce a regularized wavefunction
\begin{eqnarray}\label{eq3.5}
\psi^{reg}_{0}
&\equiv& \phi_{0}(x,R(t))e^{\textit{i}\epsilon\theta(x,R(t))}e^{-\frac{\textit{i}}{\hbar}\int^{t}_{0}E(R(t'))dt'}
\nonumber\\
&\equiv& \phi_0^{reg} (x,R(t))e^{-\frac{\textit{i}}{\hbar}\int^{t}_{0}E(R(t'))dt'}
\end{eqnarray}
and regularized potential
\begin{eqnarray}\label{eq3.6}
V^{reg}_{0}\equiv V_{0}(x,R(t))+\epsilon\tilde{V}(x,R(t)).
\end{eqnarray}
Regularized wavefunction $\psi^{reg}_{0}$ should satisfy the TDSE for regularized system,
\begin{eqnarray}\label{eq3.7}
\textit{i}\hbar\frac{\partial\psi_{0}^{reg}}{\partial t}=-\frac{\hbar^{2}}{2m}\partial_x^{2}\psi^{reg}_{0}+V^{reg}_{0}\psi^{reg}_{0},
\end{eqnarray}
up to the order of $\epsilon$. 

The potential $\tilde{V}$ is detirmined as
\begin{eqnarray}\label{2.10}
\tilde{V}=-\hbar \cdot\text{Im}\Biggl[\frac{\partial\phi_0}{\partial R}/\phi_0\Biggr]-\frac{\hbar^2}{m}\cdot\text{Im}\Biggl[\frac{\nabla\phi_0}{\phi_0}\Biggr]\cdot\nabla \theta.
\end{eqnarray}
Rewriting $\phi_{0}(x,R(t))$ in terms of the real positive amplitude $\overline{\phi}_{0}(x,R(t))$ and phase $\eta(x,R(t))$ as
\begin{eqnarray}\label{2.8}
\phi_{0}(x,R(t))=\bar{\phi}_{0}(x,R(t))e^{\textit{i}\eta(x,R(t))},
\end{eqnarray}
we see $\theta$ to satisfy
\begin{eqnarray}\label{2.9}
\partial_x(\bar{\phi}_{0}^2\partial_x\theta) =-\frac{m}{\hbar}\partial_{R}\bar{\phi}_{0}^2.
\end{eqnarray}
Integrating Eq.~(\ref{2.9}) over $x$, we have
\begin{equation}\label{2.11}
\partial_x\theta=-\frac{m}{\hbar}\frac{1}{\bar{\phi}_0^2}\int^x \partial_R\bar{\phi}_0^2 dx', 
\end{equation}
which is the core equation of the regularization procedure. 
We shall now accelerate the quasi-adiabatic dynamics of $\psi_{0}^{reg}$ in Eq.~(\ref{eq3.5}) by applying the external driving potential (fast forward potential).
For this purpose we introduce the fast-forward version 
of $\psi_{0}^{reg}$ as
\begin{eqnarray}
\label{3.13}
\psi_{FF}&=\bar{\phi}_{0}(x,R(\Lambda(t)))
e^{\textit{i}\eta(x,R(\Lambda(t)))}e^{\textit{i}v(t)\theta(x,R(\Lambda(t)))}\nonumber\\
&\times e^{-\frac{i}{\hbar}\int_0^t E(R(\Lambda(s)))ds}.
\end{eqnarray}
For accelerated system control parameter $R$, now be can rewritten as
\begin{equation}\label{rv2-1}
R(\Lambda(t))=R_0+\epsilon \Lambda(t),
\end{equation}
where $\Lambda(t)$ is the future or advanced time 
\begin{eqnarray}\label{1.3}
\Lambda(t)=\int_0^t\mathrm{\alpha(t')}\,\mathrm{d}t'.
\end{eqnarray}
The wave function of fast forward state given by Eq.~(\ref{3.13}) satisfies TDSE with a fast-forward Hamiltonian $H_{FF}$:
\begin{eqnarray}
\label{3.15}
\textit{i}\hbar\frac{\partial \psi_{FF}}{\partial t}=H_{FF} \psi_{FF}\equiv \left(-\frac{\hbar^{2}}{2m}\partial_x^2+V_{0}
+ V_{FF}\right)\psi_{FF}.\nonumber\\
\end{eqnarray}
Here $V_0=V_0(x, R(\Lambda(t)))$ and $V_{FF}$ is given by
\begin{eqnarray}
\label{3.14}
V_{FF}&=&-\frac{\hbar^{2}}{m}v(t)\partial_x\theta\cdot\partial_x\eta-\frac{\hbar^{2}}{2m}(v(t))^{2}(\partial_x\theta)^{2}\nonumber\\
&&-\hbar v(t)\partial_{R}\eta-\hbar\dot{v}(t)\theta-\hbar(v(t))^{2}\partial_{R}\theta.\nonumber\\
\end{eqnarray}
\section{Application to adiabatic dynamical confinement}
Masuda-Nakamura's fast forward protocol presented in the previous section can be applied to the simplest time-dependent system such as one-dimensional quantum box with a moving wall and one-dimensional harmonic oscillator with time-dependent frequency, evolving in the adiabatic regime. The main result of the solution of such a task should be analytically or numerically calculated wave function $\Psi_{FF}$ of the fast forwarded system and fast forwarding (driving) potential, $V_{FF}$.
\subsection{Time-dependent harmonic oscillator}
Consider first harmonic oscillator with time-dependent frequency. The evolution of such system is described in terms of the following non stationary Schr\"odinger equation:
\begin{equation}\label{leftWF}
\textit{i}\hbar\frac{\partial}{\partial t}\psi_{0}(x,R(t))= -\frac{\hbar^{2}}{2m}\partial_x^{2}\psi_{0}(x,R(t))+\frac{1}{2}m\omega^{2}(R(t))x^{2}\psi_{0}(x,R(t)),
\end{equation}
where time dependence of the frequency  $\omega(t)$ is caused by adiabatically changing parameter $R(t)$ defined as $R(t)=\sqrt{\frac{1}{\omega(t)}}$. For adiabatic regime of evolution, the  eigenvalue problem can be written in terms of the following Schr\"odinger equation:
\begin{eqnarray}\label{H-E}
H_{0}(x,R)\phi=E(R)\phi,
\end{eqnarray}
that gives the eigenvalue and the eigenstate as 
\begin{eqnarray}\label{3.5}
E_{n}&=&\left( n+\frac{1}{2}\right) \hbar\omega(R),\nonumber\\
\phi_{n}&=&\left( \frac{m\omega(R)}{\pi\hbar}\right) ^\frac{1}{4}\frac{1}{(2^n n!) ^\frac{1}{2}}\nonumber\\
&\times& e^{-\frac{m\omega(R)}{2\hbar}x^{2}} H_{n}\left(\left(\frac{m\omega(R)}{\hbar} \right)^\frac{1}{2}x \right)
\end{eqnarray}
with $n=0,1,2,\cdots$ . Here  $ H_{n}(\cdot)$ are Hermite polynomials.

Fast forward state and fast forward potential for such system can be calculated analytically and given by (see Ref.~\ci{JM} for details):
\begin{eqnarray}\label{Psi-F}
\psi_{FF}&=&\phi_{n}(x,R(\Lambda(t))e^{i\frac{m}{2\hbar}\frac{\dot{R}}{R}x^2} e^{-\left( n+\frac{1}{2}\right) i \int_{0}^{t}\omega(R(\Lambda(t')))dt'}\nonumber\equiv  \langle x|n \rangle
\end{eqnarray}
and
\begin{eqnarray}
\label{3.21}
V_{FF}=-\frac{m\ddot{R}}{2R}x^2.
\end{eqnarray}

\subsection{Time-dependent quantum box}
Now let us investigate 1D quantum box with a moving wall.  The dynamics of a particle is governed by 
\begin{eqnarray}\label{PH}
i\hbar\frac{\partial\psi}{\partial t}=H_{0}\psi=-\frac{\hbar^2}{2m}\partial_{x}^{2}\psi
\end{eqnarray}
with the time-dependent box boundary conditions as $\psi(x=0,t)=0$ and $\psi(x=L(t),t)=0$. 
$L(t)$ is assumed to change adiabatically as $L(t)=L_{0}+\epsilon t$. Length of the wall $L(t)$ is to be assumed as control parameter of the confinement. 

The adiabatic eigenvalue problem related to Eq.~(\ref{PH}) gives eigenvalues and eigenstates as
\begin{eqnarray}\label{EP}
E_{n}&=&\frac{\hbar^2}{2m}\left( \frac{\pi n}{L(t)}\right)^2,\nonumber\\
\phi_{n}&=&\sqrt{\frac{2}{L(t)}}\sin\left( \frac{\pi n}{L(t)}x\right) . 
\end{eqnarray}
The phase $\theta$ which the regularized state acquires is given using the formula in Eq.~(\ref{2.11}), as 
\begin{eqnarray}\label{theta}
\partial_{x}\theta&=&-\frac{m}{\hbar}\frac{1}{\phi_{n}^{2}}\partial_{L}\int_{0}^{x}\phi_{n}^{2}\mathrm{d}x=\frac{m}{\hbar}\frac{x}{L(t)},\nonumber\\\theta&=&\frac{m}{2\hbar}\frac{x^2}{L(t)}.
\end{eqnarray}
Thanks to the real nature of $\phi_{n}$, we find $\eta=0$ in Eq.~(\ref{2.10}) and see that $\tilde{V}$ is vanishing.

Now we apply the fast forward scheme in Section~\ref{sec2}, this will be done by changing time $t$ by future time $\Lambda(t)$ in control parameter $L(t)$.  By taking the asymptotic limit  ( $\epsilon\rightarrow 0, \bar{\alpha}\rightarrow \infty$ with $\epsilon\alpha=v(t)$ ), the fast forward state can be written as:
\begin{equation}\label{psi-ff}
\psi_{FF}=\phi_{n}\left(x,L(\Lambda(t))\right)e^{i\frac{m\dot{L}(\Lambda(t))}{2\hbar L(\Lambda(t))}x^2} e^{-i\frac{\hbar}{2m}\left(\pi n \right)^2\int_{0}^{t}\frac{\mathrm{d}t'}{L^2(\Lambda(t'))} },
\end{equation}
where $L\left( \Lambda(t)\right) =L_{0}+\int_{0}^{t}v(t')\mathrm{d}t'$
with the time scaling factor $v(t)$.
From Eq.~(\ref{3.14}),  the fast forward potential is given by
\begin{eqnarray}
\label{6.6}
V_{FF}=-\frac{m}{2}\frac{\ddot{L}(\Lambda(t))}{L(\Lambda(t))}x^2.
\end{eqnarray}
In the next section we compute energy cost needed for realization of the above models, i.e. for fast forwarding of the quantum evolution in time-dependent box and time-dependent harmonic oscillator.

\section{Energy cost needed for fast forwarding of the evolution of a quantum system }
The practical application of the above (or any other) fast forward evolution prescription is closely related to the question that, how much energy one needs to use to apply the prescription. In other words, the effectiveness of the fast forward protocol depends on the cost of energy to be spent: as smaller the energy cost as effective the protocol. Here we consider the problem for estimation of the energy cost needed for application of the Masuda-Nakamura's fast forward protocol.
According to the Ref.~\ci{cost1} energy cost to be paid for a given fast forward protocol is determined as:
\be 
C=\frac{1}{T_{FF}}\int\limits^{T_{FF}}_{0}||H|| dt,
\ee
where $||\hat{A}||$ denotes Frobenius norm and defined as $||\hat{A}||=\sqrt{\rm{Tr}\left[\hat{A}^\dagger \hat{A}\right]} $  and $H$ is total Hamiltonian of the system, which is given by
$$
H=H_0+V_{FF},
$$
where $H_0$ is the Hamiltonian of the standard system (to be fast forwarded) and $V_{FF}$ is the fast forward potential given by Eq.~\re{3.14}. 
Here we estimate the energy cost for two systems adiabatically evolving under the dynamical confinement. Namely, we consider the above described time-dependent harmonic oscillator and quantum box with a moving wall. Let us start from time-dependent harmonic oscillator given by Eq.~\re{leftWF}. 

According to \ci{cost2} the energy cost needed to fast forward a quantum system with an unbound (discrete) spectrum (which is the case for our system) can be rewritten as
\begin{equation}
    C_{FF}=\frac{1}{T_{FF}}\int\limits_{0}^{T_{FF}}\bar{U}dt
\end{equation}
where $T_{FF}$ shortened or fast-forward time and $U$ is the internal energy given by
\begin{equation}
\bar{U}={\rm{Tr}}(\rho \hat{H}_{FF}),  
\end{equation}
with density matrix $\rho$, which is defined as follows:
\begin{equation}\label{rho-expand}
\rho\left(t\right)=\sum^{\infty}_{n=0}\left|n\right\rangle f_{n}\left\langle n\right|,
\end{equation}
where $\left\{\left|n\right\rangle\right\}$ is the exact solution of TDSE in Eq.~(\ref{3.15}) and  $f_{n}$ is the Fermi-Dirac distribution which, takes
$$
f_{n}=\frac{1}{e^{\beta\left(E_{n}(L(\Lambda(t)))-\mu\right)}+1}.
$$
For time-dependent harmonic oscillator the expression of $U$ can be written as (see Ref.~\ci{JM} for details): 
\begin{eqnarray}
\bar{U}&=&A\left(\frac{\hbar^2}{4mL^2}-\frac{m}{8} L\ddot{L}+\frac{m}{8} \dot{L}^2\right)
\end{eqnarray}
with
$$
A=N^2\left[1+\frac{4\pi^2}{3}L_{0}^2\left(\frac{mkT}{\hbar^2}\right)^2 \left(\frac{N}{L_{0}}\right)^{-2}+\cdots\right]
$$
and
\begin{equation}
L=L_0+\bar{v}\left( \frac{1}{2}\frac{t^2}{T_{FF}}-\frac{1}{3}\frac{t^3}{T_{FF}^2}\right),
\lab{case1}
\end{equation}
\begin{equation}
\label{vel}
\dot{L}=\bar{v}\left(\frac{t}{T_{FF}}- \frac{t^2}{T^2_{FF}}\right).
\end{equation}
Thus for the cost we have
$$
    C_{FF}=\frac{1}{T_{FF}}\int\limits_{0}^{T_{FF}}\bar{U}dt
$$
$$
=\frac{A\hbar^2}{4m T_{FF}}\int\limits_{0}^{T_{FF}}\frac{1}{L^2}dt-\frac{mA}{8T_{FF}}\int\limits_{0}^{T_{FF}}( L\ddot{L}- \dot{L}^2)dt
$$
$$
=\frac{A\hbar^2}{4m T_{FF}}\int\limits_{0}^{T_{FF}}\frac{1}{L^2}dt+\frac{mA}{8T_{FF}}\int\limits_{0}^{T_{FF}}( -\frac{d}{dt}(\dot{L}L)+2 \dot{L}^2)dt
$$
$$
=\frac{A\hbar^2}{4m T_{FF}}\int\limits_{0}^{T_{FF}}\frac{1}{L^2}dt-\frac{mA}{4T_{FF}}\int\limits_{0}^{T_{FF}}\dot{L}^2dt,
$$
with use of Eq.~(\ref{vel}) we have
$$
mA\frac{1}{4T_{FF}}\int\limits_{0}^{T_{FF}}\dot{L}^2dt=mA\frac{1}{4T_{FF}}\int\limits_{0}^{T_{FF}}\bar{v}^2\left(\frac{t}{T_{FF}}- \frac{t^2}{T^2_{FF}}\right)^2dt
$$
$$
=mA\frac{\bar{v}^2}{4T_{FF}}\left(\frac{1}{3}T_{FF}-\frac{1}{2}T_{FF}+\frac{1}{5}T_{FF}\right)=\frac{mA}{120}\bar{v}^2,
$$
then
\begin{equation}
    C_{FF}=\frac{1}{T_{FF}}\int\limits_{0}^{T_{FF}}\bar{U}dt=\frac{A\hbar^2}{4m T_{FF}}\int\limits_{0}^{T_{FF}}\frac{1}{L^2}dt+\frac{mA}{120}\bar{v}^2.
\end{equation}

\begin{figure}[t!]
\includegraphics[width=16cm]{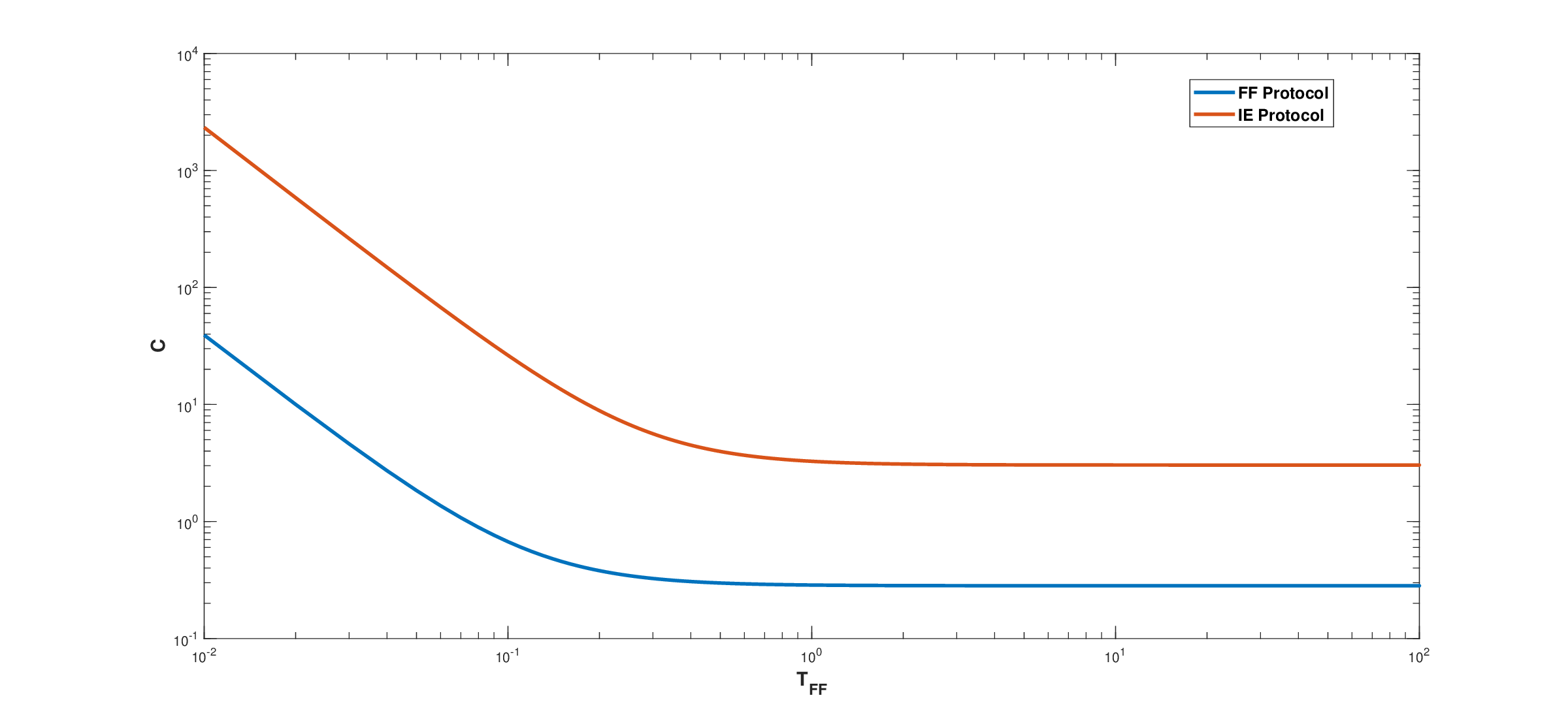}
\caption{Energy cost for the fast forward protocol (blue) with control parameter $L$ given by Eq.~(\ref{case1}) and inverse engineering protocol (red) with initial frequency $\omega_0=1$ and final frequency $\omega_F=10$.}
\label{fig::1}
\end{figure}

In the case of trigonometric function of control parameter $L(t)$
\begin{equation}
    L(t)=L_0+\bar{v}\left(t-\frac{T_{FF}}{2\pi}\sin{\frac{2\pi}{T_{FF}}t}\right),
    \lab{case2}
\end{equation}
\begin{equation}
    \dot{L}=\bar{v}\left(1-\cos{\frac{2\pi}{T_{FF}}t}\right),
\end{equation}
we will have
$$
    C_{FF}=\frac{1}{T_{FF}}\int\limits_{0}^{T_{FF}}\bar{U}dt=
$$
\begin{equation}
    =\frac{A\hbar^2}{4m T_{FF}}\int\limits_{0}^{T_{FF}}\frac{1}{L^2}dt+\frac{3Am}{8}\bar{v}^2.
\end{equation}
\begin{figure}[t!]

\includegraphics[width=16cm]{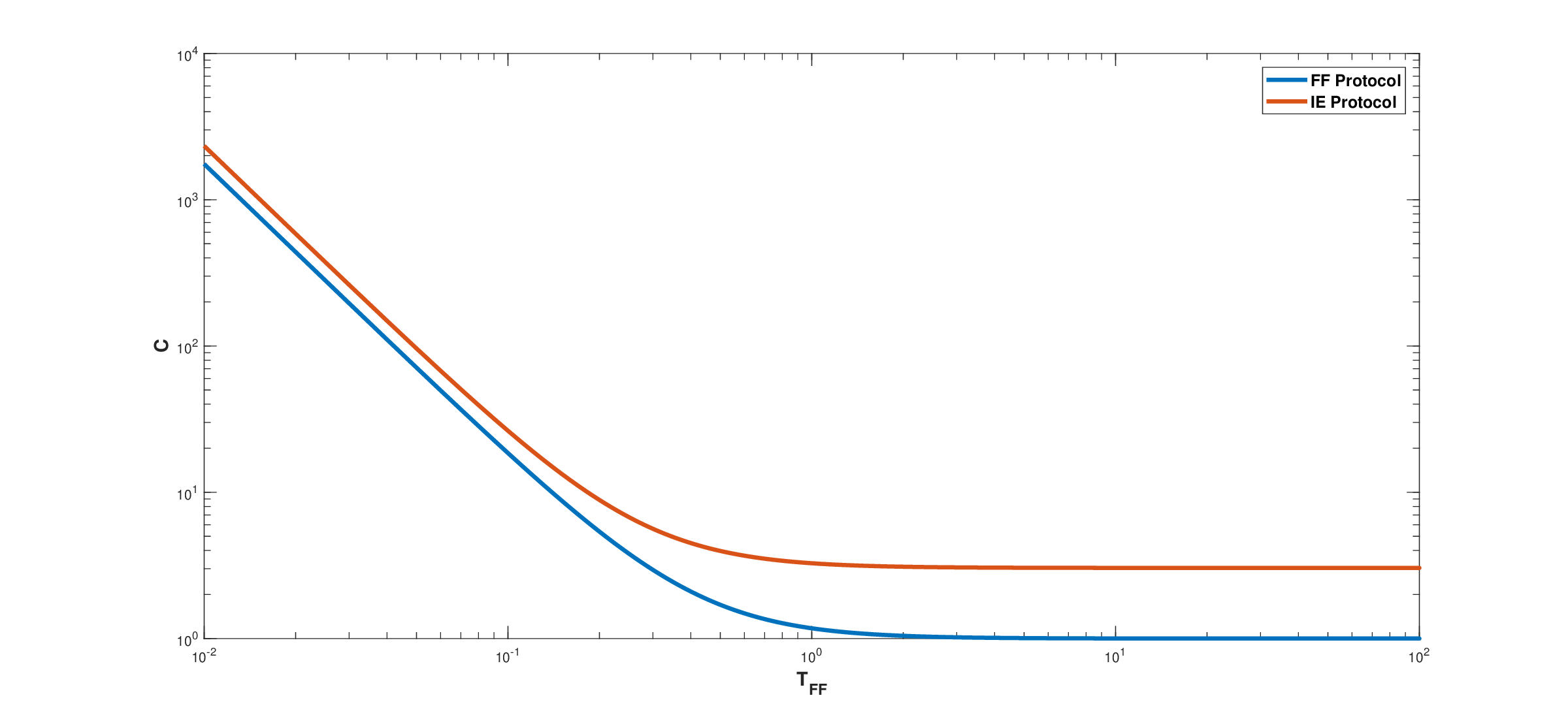}
\caption{Energy cost for the fast forward protocol (blue) with control parameter $L$ given by Eq.(\ref{case1}) and inverse engineering protocol (red) with initial frequency $\omega_0 = 1$ and final frequency $\omega_F = 10$.}
\label{fig::2}
\end{figure}

Now we will do similar calculations for quantum box with moving wall. Energy $\bar U$ for accelerated quantum box with moving wall reads as follows (for details see Ref.~\ci{JM}):
\begin{eqnarray}\label{HD-e-low}
\bar U&=&\frac{\pi^2 \hbar^2}{24m}\frac{N^3}{L^2}\Biggl[1+\frac{24}{\pi^2}\left(\frac{mkT}{\hbar^2}\right)^2\Biggl(\frac{N}{L}\Biggr)^{-4}+\cdots\Biggr]\nonumber\\
&-&\frac{N}{6}(mL\ddot{L}-m\dot{L}^2) \nonumber\\
&\times&\Biggl[1+\frac{6}{\pi^2}\frac{1}{N^2}\left(1+\frac{16}{3\pi^2}\left(\frac{mkT}{\hbar^2}\right)^2\Biggl(\frac{N}{L}\Biggr)^{-4}+\cdots \right)\Biggr].
 \nonumber\\
\end{eqnarray}
Here we also considered two cases, the first is when control parameter $L(t)$ is given as polynomial function, the second is when  $L(t)$ is trigonometric function. As for polynomial $L(t)$:
\begin{equation}
L=L_0+\bar{v}\left( \frac{1}{2}\frac{t^2}{T_{FF}}-\frac{1}{3}\frac{t^3}{T_{FF}^2}\right)
\lab{case3}
\end{equation}
energetic cost is given by 
\begin{equation}
    C_{FF}=\frac{1}{T_{FF}}\int\limits_{0}^{T_{FF}}\bar{U}dt=B_1\frac{1}{24 T_{FF}}\int\limits_{0}^{T_{FF}}\frac{1}{L^2}+B_2\frac{\bar{v}^2}{90}
\end{equation}
where $B_1$ and $B_2$ are constant terms
\begin{equation}
B_1= \frac{\pi^2 \hbar^2N^3}{24m}\Biggl[1+\frac{24}{\pi^2}\left(\frac{mkT}{\hbar^2}\right)^2\Biggl(\frac{N}{L}\Biggr)^{-4}+\cdots\Biggr],   
\end{equation}
\begin{equation}
B_2= \frac{mN}{6}\Biggl[1+\frac{16}{3\pi^2}\left(\frac{mkT}{\hbar^2}\right)^2\Biggl(\frac{N}{L}\Biggr)^{-4}+\cdots\Biggr].  
\end{equation}
For trigonometric control parameter $L(t)$
\begin{equation}
    L(t)=L_0+\bar{v}\left(t-\frac{T_{FF}}{2\pi}\sin{\frac{2\pi}{T_{FF}}t}\right),
    \lab{case4}
\end{equation}
for the energetic cost we obtain the next expression
\begin{equation}
    C_{FF}= \frac{1}{T_{FF}}\int\limits_{0}^{T_{FF}}\bar{U}dt=B_1\frac{1}{24 T_{FF}}\int\limits_{0}^{T_{FF}}\frac{1}{L^2}+B_2\frac{\bar{v}^2}{2}.
\end{equation}

\begin{figure}[t!]
\includegraphics[width=16cm]{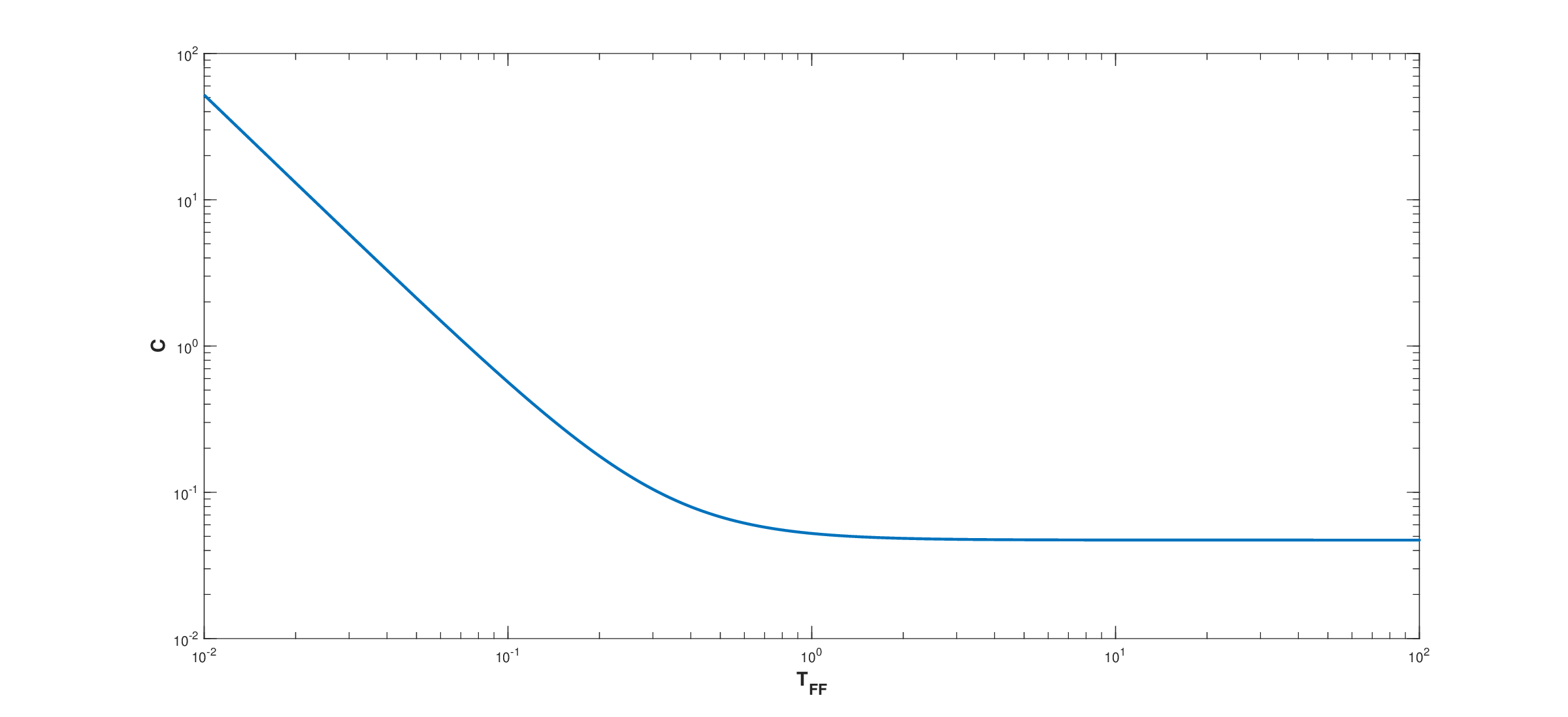}
\caption{Energy cost of the fast forward protocol with $L$ given by Eq.~(\ref{case3}) for the parameters $L_0 = 1$ and  $L_F = 10$.}
\label{fig::3}
\end{figure}

\begin{figure}[t!]
\includegraphics[width=16cm]{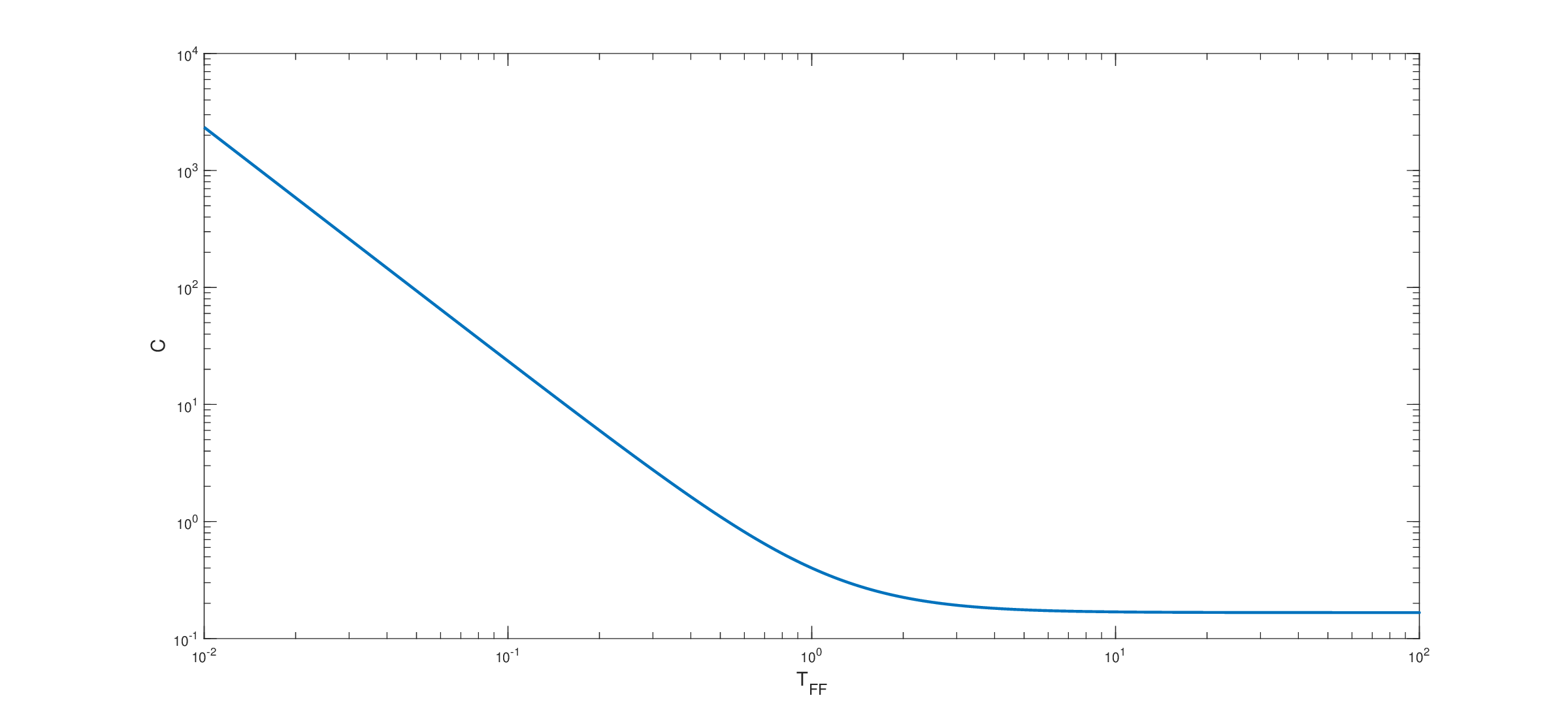}
\caption{Energy cost of the fast forward protocol with $L$ given by Eq.~(\ref{case4}) for the parameters $L_0 = 1$ and  $L_F = 10$.}
\label{fig::4}
\end{figure}

Fig.~\ref{fig::1} compares plots of the energy cost as function of time for Masuda-Nakamura's (blue line) and  the inverse engineering (IE) (red line) (see, the Ref.~\ci{cost2} for details) fast forward  protocols for time-dependent harmonic oscillator. Energy cost for IE protocol is obtained numerically by using following IE Hamiltonian~\ci{cost2}  :
\be
\langle H_{IE}\rangle=\frac{1}{2}\Big[\frac{\dot{ b}^2(t)}{2\omega_0}+\frac{\omega^2(t)b^2(t)}{2\omega_0}+\frac{\omega_0}{2b^2(t)} \Big]\coth\Big(\frac{\beta\omega_0}{2}\Big),
\ee
where $b(t)$ is the dimensionless function satisfying Ermakov equation:
\be
\ddot{b}(t)+\omega^2(t)b(t)=\omega_0/b^3(t).
\ee
The calculations are done for the control parameter, $L(t)$ given by Eq.~\re{case1}. As the plot shows, the cost for Masuda-Nakamura protocol is much smaller than that for the inverse engineering one and the curves are almost parallel to each other. Fig.~\ref{fig::2} presents similar plots for the form of $L(t)$ given by Eq.~\re{case2}. As it can be seen, the costs are completely different than that in Fig.~\ref{fig::1}, both qualitatively and quantitatively, i.e. the cost for Masuda-Nakamura's fast forward protocol in Fig.~\ref{fig::1} is much smaller than that in Fig.~\ref{fig::2}. In addition, at the initial time the difference between the costs are much smaller than that at longer times. In Fig.~\ref{fig::3} time-dependence of the energy cost for Masuda-Nakamura protocol is plotted for a quantum box with a moving wall for the control parameter given by Eq.~\re{case1}. Fig.~\ref{fig::4} presents similar plot for the case, when $L(t)$ is given by Eq.~\re{case2}. Comparing plots of the costs presented in Figs.\ref{fig::3} and \ref{fig::4} with those in Figs.~\ref{fig::1} and \ref{fig::2} one can conclude that the costs for time-dependent harmonic oscillator and quantum box with a moving wall within the Masuda-Nakamura's fast forward protocol are almost equal for the same systems (provided they are estimated for the same control parameter). 

\section{Conclusion}
In this paper we proposed two models where Masuda-Nakamura's fast forward protocol can be applied in the adiabatic regime  and  the energy cost for the realization of such protocol can be computed. In particular, Masuda-Nakamura's method for fast-forward evolution, \ci{mas1,mas2} is applied for the acceleration of the evolution of the time-dependent box with a slowly moving wall and harmonic oscillator with slowly varying time-dependent frequency. Done quantitative comparison energetic cost of Masuda-Nakamura protocol with inverse engineering protocol. In particular, the plots of energy cost for Masuda-Nakamura and inverse engineering protocols in Figs.~\ref{fig::1} and \ref{fig::2} shows that Masuda-Nakamura protocol requires less cost than the inverse-engineering protocol. The results obtained in this paper can be used for further development of energy-efficient and resource-saving low-dimensional quantum devices.

\newpage

\section*{Acknowledgement}
This work is supported by the grant of the Agency of Innovative Development of the Republic of
Uzbekistan, World Bank Project "Modernizing Uzbekistan National Innovation System" (Ref. Nr. REP-
05032022/235).

\end{document}